\documentclass[aps,prd,a4paper,nofootinbib,showpacs,showkeys,
preprintnumbers,superscriptaddress,twocolumn]{revtex4}

\usepackage{amsmath}
\usepackage{amssymb}
\usepackage{bm}
\usepackage{graphicx}

\begin{document}

\title{Thermodynamics of the Quark-Gluon Plasma in Terms of Quasiparticles and Polyakov Line Condensates}

\author{Lucia Oliva}
\affiliation{Department of Physics and Astronomy, University of Catania, Via S. Sofia 64, I-95125 Catania}
\affiliation{INFN-Laboratori Nazionali del Sud, Via S. Sofia 62, I-95123 Catania, Italy}

\author{Paolo Castorina}
\affiliation{Department of Physics and Astronomy, University of Catania, Via S. Sofia 64, I-95125 Catania}
\affiliation{INFN-CT, Via S. Sofia 64, I-95123 Catania, Italy}

\author{Vincenzo Greco}
\affiliation{Department of Physics and Astronomy, University of Catania, Via S. Sofia 64, I-95125 Catania}
\affiliation{INFN-Laboratori Nazionali del Sud, Via S. Sofia 62, I-95123 Catania, Italy}

\author{Marco Ruggieri}
\affiliation{Department of Physics and Astronomy, University of Catania, Via S. Sofia 64, I-95125 Catania}


\begin{abstract}
We study a model of quark-gluon plasma of 2+1 flavors Quantum Chromodynamics 
in terms of quasiparticles propagating in a condensate of Polyakov loops.  
The Polyakov loop is coupled to quasiparticles by means of a gas-like effective potential.
This study is useful to identify the effective degrees of freedom propagating in the medium above the
critical temperature.
Our finding is that a dominant part of the phase transition dynamics
is accounted for by the Polyakov loop, 
hence the thermodynamics can be described without the need for 
rapidly increasing quasiparticle masses as $T \rightarrow T_c$, 
at variance respect to standard quasiparticle models.
\end{abstract}


\maketitle

{\em Introduction.} The interest in understanding the
thermodynamic properties of the strong interaction theory (QCD)
has noticeably increased in the recent years,
mainly thanks to improvement of computer facilities which allow to perform
lattice simulations of QCD and other Yang-Mills theories, as well as to the 
possibility to create in laboratories extremely hot environments
by means of heavy ion collisions.
Lattice simulations of the pure gauge $SU(3)$ 
theory have shown a deconfinenemt phase transition
at $T= T_c \approx 270$ MeV~\cite{Datta:2010sq,Boyd:1996bx,Boyd:1995zg,Borsanyi:2011zm}. 
Introducing dynamical quarks the phase transition turns to a smooth 
crossover~\cite{Aoki:2006br,Aoki:2009sc,Borsanyi:2010cj,Bazavov:2009zn,Cheng:2009be}. 
In this case it is not possible to define rigorously a transition temperature,
because of the absence of a true phase transition; nevertheless a transition region
centered on a pseudocritical temperature, which we still denote by $T_c$,
can be identified as the one where the thermodynamic quantities have the
maximum variation; this definition leads to $T_c\approx 155$ MeV
in the case of QCD with $u$, $d$ and $s$ quarks. 

For what concerns the high temperature phase 
it is customary to identify the system above $T_c$ as a plasma
of quarks and gluons.
However, above the critical temperature 
the in-medium interactions are nonperturbative,
and this
makes the identification of the correct degrees of freedom of the quark-gluon plasma, in proximity
as well as well beyond the critical temperature, a very complicated task.
Resummation schemes have been proposed, based for example on the Hard Thermal Loop (HTL) approach~\cite{HTL1,HTL3,HTL4,HTL7,Andersen_HTLpt}. At very high temperature the HTL 
approach motivates and justifies a picture of weakly interacting quasi-particles, as 
determined by the HTL propagators. 

This quasiparticle description
has been assumed to be valid also in the case
of $T\approx T_c$~\cite{Gorenstein:1995vm,PKPS96, Levai:1997yx,bluhm-qpm,
Meisinger:2003id,Castorina:2011ra,Giacosa:2010vz,Peshier:2005pp,redlich,
Plumari:2011mk,Filinov:2012pt,Castorina:2007qv,Castorina:2011ja,Brau:2009mp,Cao:2012qa,Bannur:2006hp}. 
In such an approach, one assumes that the quark-gluon plasma description of the
deconfinement phase, with propagating transverse gluons and quarks, is still valid;
the strong interaction in this non perturbative regime
is taken into account through a temperature-dependent mass for the 
propagating degrees of freedom.
Within this framework one usually assumes a dependence 
of the quasiparticle masses on the temperature, leaving few free parameters 
which are then fixed by fitting the thermodynamical data of lattice simulations. 
This description of the quark-gluon plasma is interesting because
it is possible to include the quasiparticle dynamics 
into a transport theory capable to directly 
simulate the expanding fireball produced in heavy ion collisions
computing the collective properties, as well as the chemical 
composition of the fireball as a function of 
time~\cite{Scardina:2012hy,Ozvenchuk:2012fn,Cassing:2009vt,Bratkovskaya:2011wp}. 

In this brief report we study an extension of the quasiparticle picture of the finite
temperature QCD medium, supporting a picture in which quark and gluon quasiparticles
propagate in, and interact with, a background Polyakov loop~\cite{Polyakov:1978vu,Susskind:1979up,Svetitsky:1982gs,Svetitsky:1985ye}, following previous studies
which within this scheme took into account only the pure glue medium~\cite{Meisinger:2003id,Sasaki:2012bi,Ruggieri:2012ny}. 
The standard quasiparticle approach accounts for the dynamics 
at the onset of deconfiment only by means of temperature dependent 
masses, which leads to diverging (or steadily increasing) masses
as $T\rightarrow T_c$.
On the other hand, it has been shown that combining a $T$-dependent quasiparticle 
mass with the Polyakov
loop dynamics results in a quite different behavior of the mass itself as $T\rightarrow T_c$
\cite{Meisinger:2003id,Sasaki:2012bi,Ruggieri:2012ny}, 
at least in the case of the pure glue system; the purpose of the present study is to show that
this regular behavior of the quasiparticle masses holds even in the case of 
QCD with dynamical flavors.

In our study we introduce an effective potential for the Polyakov loop,
and couple the latter in a similar manner to what is done
within the Polyakov extended Nambu-Jona Lasinio 
model~\cite{Fukushima:2003fw,Ratti:2005jh}. We follow the
formalism of~\cite{Sasaki:2012bi}, extending it to the case of QCD
with dynamical quarks. Our main finding is that also with dynamical quarks 
the Polyakov loop background is sufficient to take into account of the
nonperturbative aspects of the QCD crossover, and quasiparticle masses
are regular as temperature approaches $T_c$.  

{\em Quasiparticle model.} As explained in the Introduction,
the main purpose of our study is to confirm that the presence of the Polyakov line condensates in the
quark-gluon plasma phase mitigates significantly the divergence of the quasiparticle masses in the crossover region.
We also find that the above result is independent on the ability to 
reproduce correctly the lattice results on the expectation value of the Polyakov loop, whenever the
latter is smaller than one in the crossover region (if the Polyakov loop was about one in the 
crossover region as well, then the model would not be different from the pure quasiparticle
ones, which predict large masses in that temperature range). 

The Polyakov loop in the representation $R$ is defined as $\ell_R = \text{Tr}L_R/d_R$
where 
\begin{equation}
L_R(\bm x) = \text{T}\exp
\left[i g \int_0^\beta T^a_R A_4^a(\tau,\bm x) d\tau\right]~,
\end{equation}
with $T^a_R$ ($a=1,\dots,N_c^2 - 1$) corresponds to the generator of the color group $SU(N_c)$
in the representation $R$, and $d_R$ corresponds to the dimension of the representation. 
In this study both the loops in the fundamental and adjoint representations
will be relevant. We follow here the approach of~\cite{Sasaki:2012bi} which
has been developed for the pure gauge theory, extending it to the case
in which dynamical quarks are also present in the thermal bath.
The thermodynamic potential $\Omega$ is given by the sum of several contributions,
\begin{equation}
\Omega = \Omega_\ell + \Omega_g + \Omega_q~,
\label{eq:Omega}
\end{equation}
where
\begin{equation}
\Omega_{\ell} = -a T\log\left(1 -6\ell_F^2 +8\ell_F^3 -3\ell_F^4\right) + c
\label{eq:Omega_l}
\end{equation}
corresponds to the pure Polyakov loop potential,
\begin{equation}
\Omega_{g} = 2T\int\frac{d^ 3p}{(2\pi)^3}\text{Tr}\log
\left(1-L_A e^{-\beta\omega_g}\right)
\label{eq:Omega_g}
\end{equation}
is the transverse gluon quasiparticles potential in the Polyakov loop background, and
\begin{equation}
\Omega_{q} = -4T\sum_f\int\frac{d^ 3p}{(2\pi)^3}\text{Tr}\log
\left(1+L_F e^{-\beta\omega_f}\right)
\label{eq:Omega_q}
\end{equation}
corresponds to the quark quasiparticles potential. In the above equations $L_F$ and $L_A$
denote the Polyakov line in the fundamental and adjoint representations respectively.
The dispersion laws for gluon quasiparticles are given by $\omega_g = \sqrt{\bm p^2 + m_g^2}$ with
\begin{equation}
m_g =\sqrt{\frac{3}{4}}g(T) T~, 
\end{equation}
where we assume 
\begin{equation}
g(T)^2=\frac{8\pi^2}{9}\frac{1}{\log\left[\left(T-w\right)/q\right]}~.
\label{eq:gT}
\end{equation}
On the same footing for the quark quasiparticles we assume $\omega_f = \sqrt{\bm p^2 + m_f^2}$ with
\begin{equation}
m_q = \sqrt{\frac{1}{3}}g(T) T
\end{equation}
and $m_f = m_q$ for $f=u,d$, while $m_s = m_0 + m_q$ with $m_0 = 95$ MeV. 

In our study we restrict ourselves to the mean field approximation which amounts to replace
$\ell_R \rightarrow\langle\ell_R\rangle$ in the thermodynamic potential; to avoid heavy notation
we denote the expectation value with $\ell_R$ from now on, unless otherwise specified.
Moreover we consider only a quark-gluon plasma at zero baryon chemical potential,
which implies $\ell_F = \bar\ell_F$. The traces in Eqs.~\eqref{eq:Omega_g} and~\eqref{eq:Omega_q} 
are easily performed in the Polyakov gauge, where the Polyakov lines are diagonal. 
In agreement with~\cite{Sasaki:2012bi} in order to simplify the calculations setup, 
we assume that the relation
\begin{equation}
(N_c^2 - 1)\ell_A = N_c^2\ell_F^2 -1~,
\end{equation}
which is valid for the actual operators, turns to a relation for the mean fields
in which $\ell_F^2 \rightarrow \langle\ell_F\rangle^2$,
which permits to express both $\Omega_g$ and $\Omega_q$ in terms of $\ell_F$. 
This approximation is known to lead to negative values of the adjoint loop~\cite{Zhang:2010kn,Kahara:2012yr}
in the low temperature phase, and to avoid this problem a matrix model on the lines
of~\cite{Abuki:2009dt} should be considered; however we have verified that in our calculations, 
which refer to the high temperature phase of QCD, the adjoint loop
is always positive. We treat $\ell_F$
as a variational parameter, imposing that at any temperature the stationarity condition
$\partial\Omega/\partial\ell_F=0$ is satisfied. 

{\em Results and discussion.}
\begin{figure}[t!]
\begin{center}
\includegraphics[width=7.5cm]{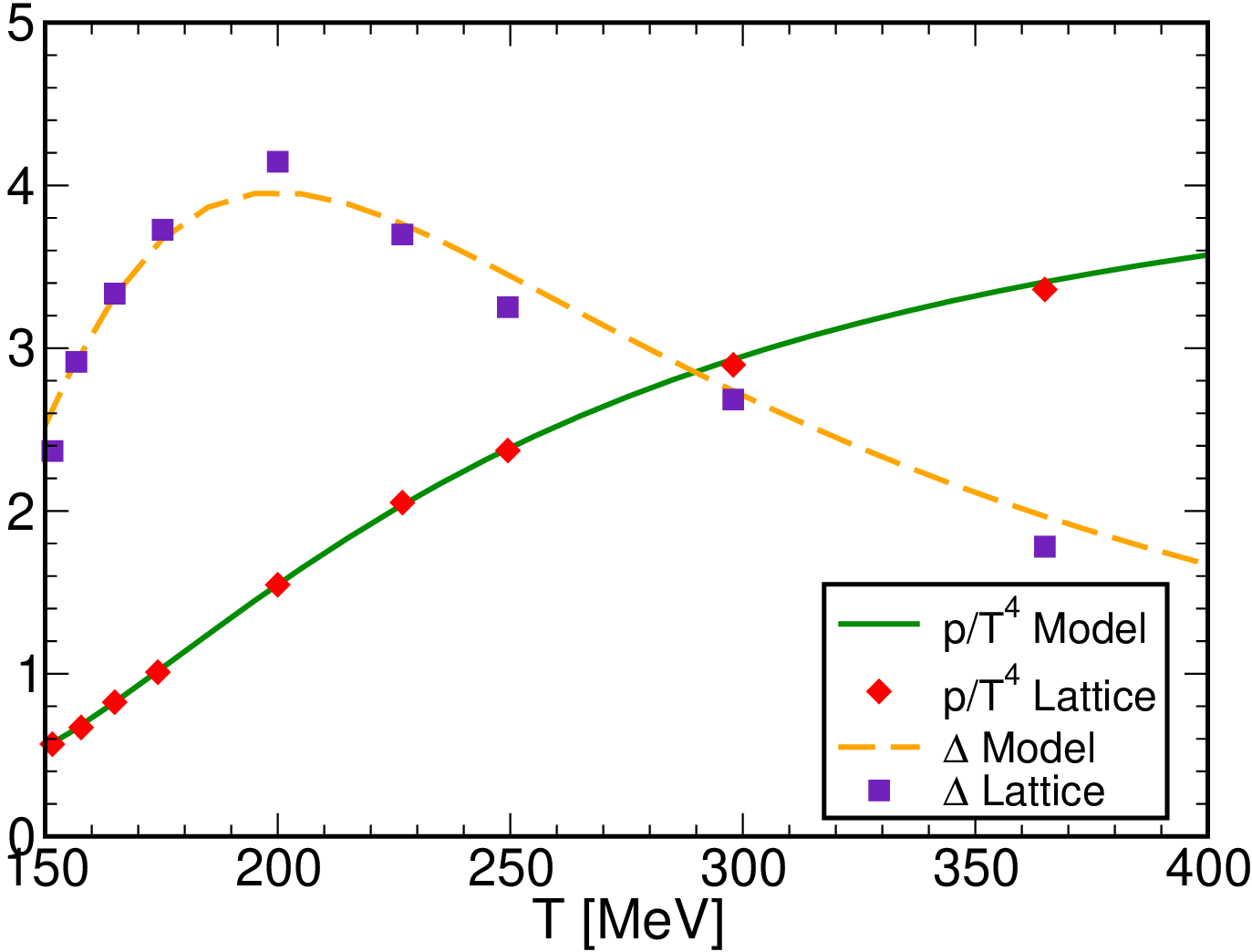}
\end{center}
\caption{\label{fig:p1}{\em Color online.} 
Pressure and interaction measure as a function of temperature
computed within the quasiparticle model. 
Lattice data~\cite{Borsanyi:2010cj} are represented by the red diamonds
and indigo squares.}
\end{figure}
In this study we have considered two different approaches. In the first one, 
which we call Model I, we fix the four free parameters
of the model by a best fit procedure of the lattice data with the computed pressure $p=-\Omega$. From the operative
point of view we proceed as follows. After fixing one set of the values of the parameters, at any temperature
we compute the numerical value of the Polyakov loop according to the stationarity condition
and then the pressure according to Eq.~\eqref{eq:Omega}; we iterate this for several values of
temperature at which lattice data are available, then computing the mean squared deviation 
of the computed pressure from the
data themselves. We repeat the procedure making a scan of the parameter space, and we finally choose
the parameter set for which we obtain the minimum value of the mean squared deviation. 
Within this procedure, the expectation value of $\ell_F$ is an output of the calculation,
since we do not impose any constraint on it besides the stationarity condition. 
This procedure leads to the values $a=(147.36\text{~ MeV})^3$, 
$c=(64.21\text{~ MeV})^4$, $w=8$ MeV and $q=5$ MeV.

In Fig.~\ref{fig:p1}, we plot the pressure as a function of temperature
obtained within the best fitting procedure described above, 
and the lattice data for the pressure. 
For completeness, in the same figure we plot also the interaction measure defined as
\begin{equation}
\Delta = \frac{\varepsilon - 3p}{T^4}~,
\end{equation}
as a function of temperature, and compare the model result with the lattice data.

\begin{figure}[t!]
\begin{center}
\includegraphics[width=7.5cm]{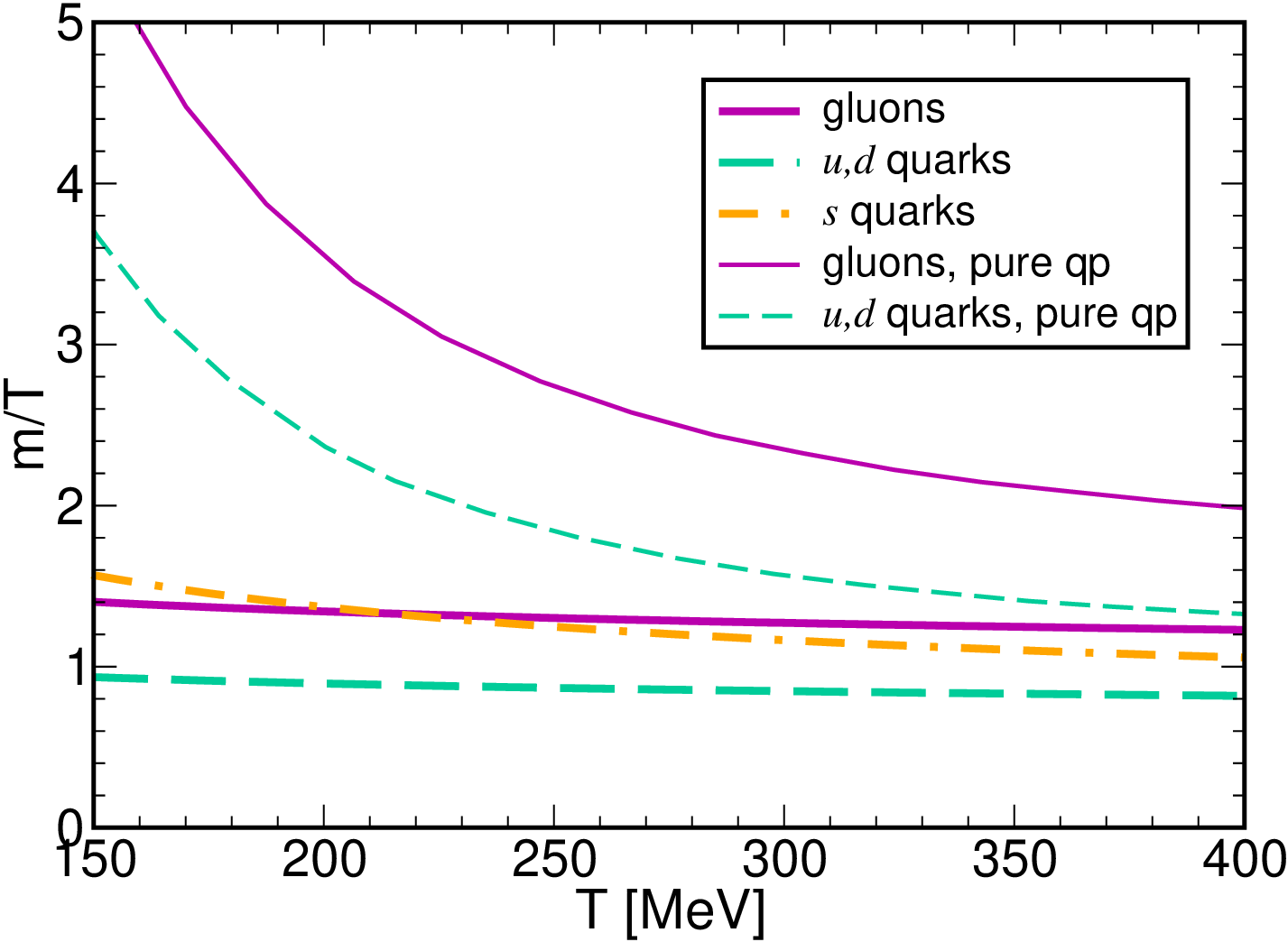}
\end{center}
\caption{\label{fig:p2}{\em Color online.} Ratio of quasiparticle masses over temperature as a function of
temperature. Pure qp denote the results for the pure quasiparticle model.}
\end{figure}

In Fig.~\ref{fig:p2}, we plot the ratio of quasiparticle masses over temperature as a function of
temperature, for the model with the Polyakov loop. For comparison we also show
by thin lines the
results obtained in the pure quasiparticle model, which is obtained by our model
setting $\ell_F=1$ in the quasiparticle thermodynamic potential and neglecting
the pure Polyakov loop potential. 
The point we stress in this brief report, which is summarized in Fig.~\ref{fig:p2}, 
is that the presence of the Polyakov loop background avoids the stiff increase of the quasiparticle 
masses as the critical region is approached from larger temperatures, which instead is a
characteristic of the model in which no Polyakov loop background is added.

In the latter model the increasing masses are understood easily since the lattice pressure
in the critical region decreases rapidly as the system is cooled down, and this can be
reproduced within the quasiparticle model only assuming large increase of the masses which
results in the suppression of the states relavant for the thermodynamics.
On the other hand, in our model it is no longer necessary that masses become larger
and larger as the critical temperature is reached from above, because the states
are suppressed statistically thanks to the coupling with the Polyakov loop. This mechanism
is similar to the statistical confinement mechanism present in the PNJL 
model~\cite{Fukushima:2008wg,Abuki:2008nm}, and is understood
as follows: the quark quasiparticles (for gluons the discussion is similar) potential
can be written as
\begin{equation}
\Omega_q =-4 T\sum_f \int\frac{d^3 p}{(2\pi)^3}
\log\left(1 + 3\ell_F x + 3\ell_F x^2 + x^3\right)~,
\label{eq:Ospieg}
\end{equation} 
where $x=e^{-\beta\omega_q}$. In the crossover region the Polyakov loop $\ell_F \ll 1$, 
which results in the suppression of the one-quark and two-quark states contributions
to the thermodynamic potential, hence suppressing quasiparticle pressure
even if $m/T$ does not become larger.

\begin{figure}[t!]
\begin{center}
\includegraphics[width=7.5cm]{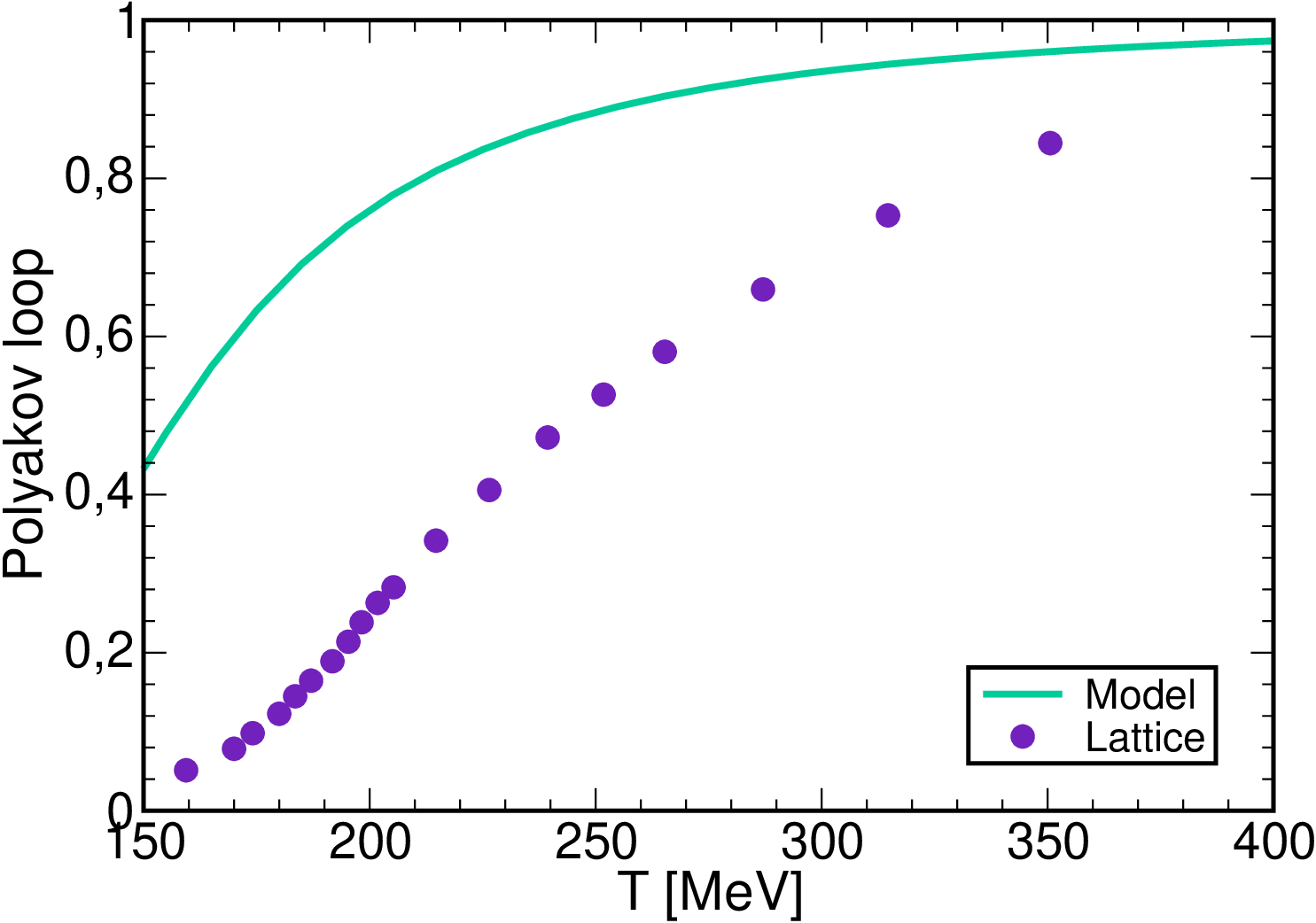}
\end{center}
\caption{\label{fig:p3}{\em Color online.} Expectation value of $\ell_F$ as a function of
temperature computed within the model, and compared with the lattice data~\cite{Bazavov:2009zn}.}
\end{figure}
In Fig.~\ref{fig:p3} we plot the model prediction for the Polyakov loop expectation value, and compare
the results with the most recent lattice data~\cite{Bazavov:2009zn}. The model prediction is obtained
solving the selfconsistent relation $\partial\Omega/\partial\ell_F = 0$. As it is evident from the figure,
the model overestimates the result for $\ell_F$. This overestimate was obtained also
in the case of the model of the pure $SU(3)$ gauge model~\cite{Ruggieri:2012ny}. 
We have not found
a consistent solution to this problem, which probably resides in the coupling of the 
Polyakov loop to the quasiparticles. However, given this discrepancy between lattice data
and model predictions about $\ell_F$, it is interesting to ask whether our results for the
quasiparticle masses are affected by the dynamical detail of $\ell_F$.  

To explore this point in more detail, we need a model in which the Polyakov loop
is in agreement with lattice data. To this end, we need to 
modify the best fit procedure above by requiring that $\ell_F$ reproduces
the lattice value at any temperature. 
We are aware this is a very rough procedure, and a more interesting study would be to 
investigate the reason of the discrepancy with the lattice results. Leaving to a 
future study a more detailed investigation, we accept here a simplicistic point of view
and just try to build up a model in which the Polyakov loop is in agreement with
the lattice data, to understand how this affects the quasiparticle masses.
We call this model as Model II.

\begin{figure}[t!]
\begin{center}
\includegraphics[width=7.5cm]{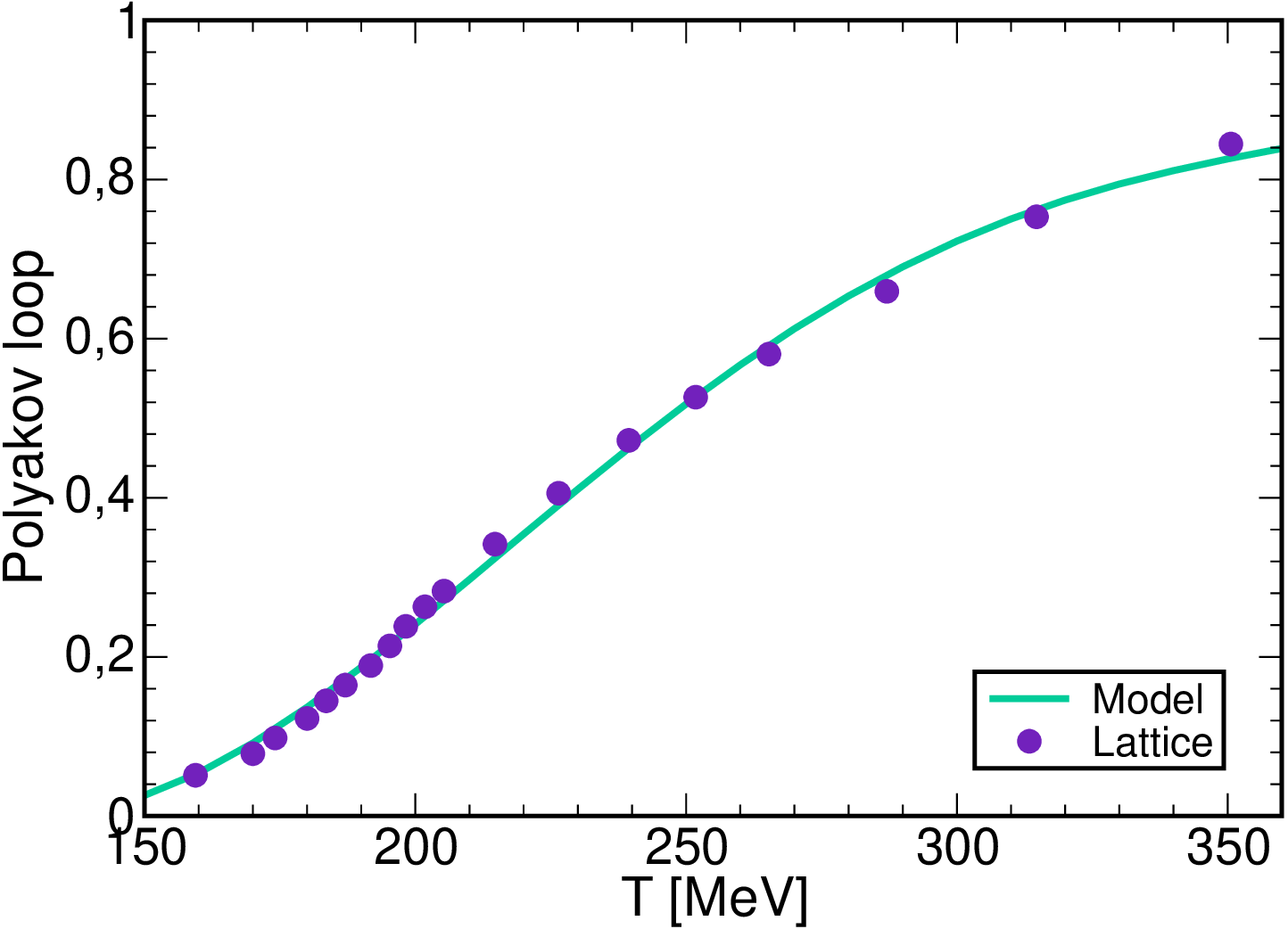}
\end{center}
\caption{\label{fig:p32}{\em Color online.} Expectation value of $\ell_F$ as a function of
temperature computed within the model, and compared with the lattice data~\cite{Bazavov:2009zn}.}
\end{figure}
In Fig.~\ref{fig:p32} we plot the expectation value of the Polyakov loop as a function of temperature
for the Model II. In order to reproduce both the pressure and the Polyakov loop lattice data 
better than we do by Model I we have replaced the parameter $c$ in Eq.~\eqref{eq:Omega_l} by a three parameters function, 
namely $c(T)=\alpha T^4 \exp(-(x-x_0)/\gamma)^2$ with $\alpha=1.62$, $\gamma=141.4$ MeV and $x_0=260.1$ MeV.

\begin{figure}[t!]
\begin{center}
\includegraphics[width=7.5cm]{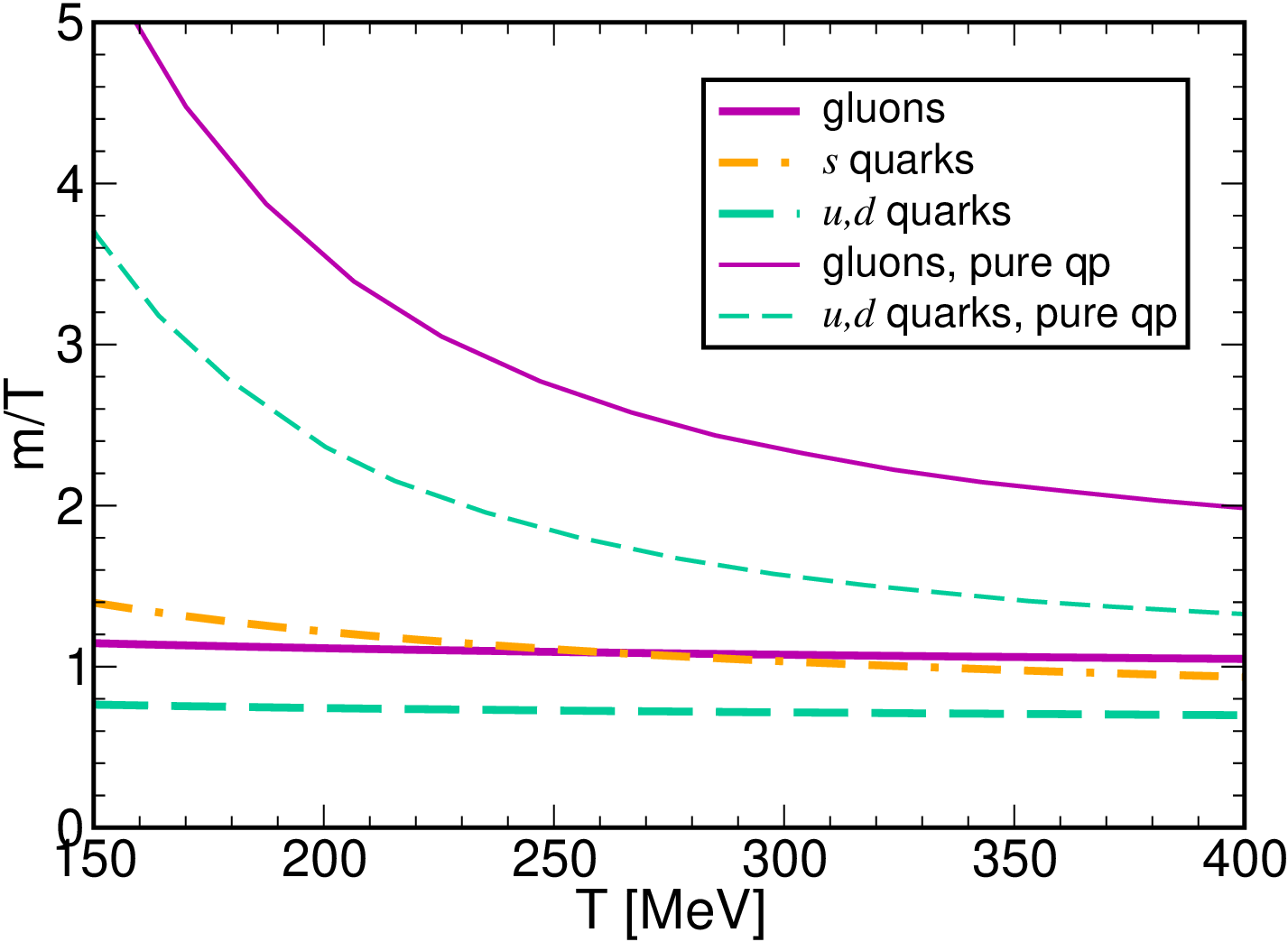}
\end{center}
\caption{\label{fig:p4}{\em Color online.} Ratio of quasiparticle masses over temperature as a function of
temperature, for the model II in which the lattice Polyakov loop is used as an input.}
\end{figure}
In Fig.~\ref{fig:p4}, we plot the ratio of quasiparticle masses over temperature as a function of
temperature for the case of model II. For this model we reproduce lattice 
data for pressure and interaction measure with the same accuracy of Model I,
therefore we do not show explicitly the data.
Line and color conventions in Fig.~\ref{fig:p4} are the same of Fig.~\ref{fig:p2}. In the fitting procedure
we require that the model reproduces both the total pressure and the lattice data on the Polyakov loop.
Once again we find that the quasiparticle masses are not rapidly increasing as the critical
region is approached, and indeed the behaviour is quite similar to Model I, see Fig.~2. 
The reason is still related to the mechanism of statistical 
confinement we discussed above: in the case $\ell_F$ is in agreement with the lattice data,
it is smaller than the one we obtained within the previous procedure and plotted in Fig.~\ref{fig:p3};
this implies that quasiparticles are even more statistically suppressed, and to reproduce
the total pressure one needs to lower the masses by about $15\%$ to allow the Boltzmann factors to compete with
the lowering of $\ell_F$.

{\em Conclusions.} In this brief report we have studied a model of quark-gluon
plasma which combines the description in terms of dynamical quasiparticles
with that of a condensate of Polyakov lines. Our main purpose has been to
discuss how the presence of the Polyakov loop background affects the 
quasiparticle masses in the critical region. We have found that the Polyakov loop
coupling to the quasiparticles helps to suppress the states in the
critical region, permitting the masses to increas not in the same
region. This behaviour is different from what is usually found
in pure quasiparticle models, where the statistical suppression of states
in the critical region can be achieved only by assuming a rapid increas
of the quasiparticle masses as the critical temperature is approached from
above.  

We have found that within our simple model, the computed expectation value
of the Polyakov loop is quite different from that computed on the lattice.
In order to understand how this discrepancy affects the result on the
quasiparticle masses, we have slightly modified our model by using
the lattice $\ell_F$ as an input. The result is summarized in Fig.~\ref{fig:p4} which shows
that the masses are only moderately affected by the different Polyakov loop
expectation value.

The straighforward step ahead is to investigate on possible different
couplings of the Polyakov loop background and the quasiparticles,
in order to build a model in which $\ell_F$ is faithful to lattice
data in a more natural way. We leave this interesting question 
to a near future project.

\end{document}